\documentclass[a4paper,12pt]{article}
\usepackage{cite}
\usepackage{graphicx}
\usepackage{amssymb}
\usepackage{amsmath}
\usepackage{amsfonts}
\usepackage{dsfont}
\usepackage{mathtools}
\usepackage{slashed}
\usepackage{amsmath}
\usepackage{caption}

\usepackage{rotating}
\usepackage{bbold,amsfonts}

\usepackage[utf8]{inputenc}
\usepackage{bm}
\usepackage{xcolor}
\usepackage{float}
\usepackage{braket}
\numberwithin{equation}{section}
\usepackage[height=8.8in,width=6.45in]{geometry}
\usepackage[font=small,labelfont=bf]{caption}
\usepackage[hidelinks]{hyperref}
\begin{document}
	\begin{titlepage}
		\vbox{
			\halign{#\hfil         \cr
			} 
		}  
		\vspace*{15mm}
		\begin{center}
			{\Large \bf 
				Holographic M5 branes in $AdS_7 \times S^4$
			}
			
			\vspace*{15mm}
			
			{\large Varun Gupta}
			\vspace*{8mm}
			
			Chennai Mathematical Institute, \\
			SIPCOT IT Park, Siruseri 603103, India \\
			
			\vskip 0.8cm
			
			
			{\small
				E-mail:  vgupta@cmi.ac.in
			}
			\vspace*{0.8cm}
		\end{center}
		
		\begin{abstract}
			
			We study classical M5 brane solutions in the probe limit in the $AdS_7 \times S^4$ background geometry that preserve the minimal amount of supersymmetry. These solutions describe the holography of codimension-2 defects in the 6d boundary dual $\mathcal{N}= (0,2)$ supersymmetric gauge theories. The general solution is described in terms of holomorphic functions that satisfy a scaling condition. We show the behavior of the world-volume of a special class of BPS solutions near the $AdS$ boundary region can be characterized by general equations, which describe it as intersections of the zeros of holomorphic functions in three complex variables with a 5-sphere. 

		\end{abstract}
		\vskip 1cm
		{
		}
	\end{titlepage}

	\setcounter{tocdepth}{2}
	\tableofcontents
	\vspace{.5cm}
	\begingroup
	\allowdisplaybreaks
	 
		\section{Introduction and summary of the results}
		Non-local operators are important observables in the Supersymmetric gauge theories in various dimensions. Line operators such as Wilson operators and 't Hooft operators are familiar examples. They serve as order parameters in the gauge theories that provide the valuable information about the phase structure in the non-perturbative regime. Similarly, Surface operators of Gukov and Witten \cite{Gukov:2006jk} with additional parameters  
		are expected to provide more information about the topological phases of gauge theories \cite{Gukov:2013zka}.
		Surface operators in the 4d supersymmetric gauge theories are two-dimensional operators which may descend from the non-local operators in the six-dimensional $(0,2)$ gauge theory, after suitable dimensional compactifications. These non-local operators in 6d theory can be either of codimension-2 or codimension-4 and they have been a focus of many investigations \cite{Drukkeretal:0320, Drukkeretal:0420,FGT:2015}. Most of these studies have been done using the constructions in M-theory with intersecting $M5$ branes or $M2$-$M5$ branes. Recently, in \cite{Drukkeretal:0320, Drukkeretal:0420} the analysis has used the probe M2 branes in the $AdS_7 \times S^4$ background to discuss a lot about the codimension-4 defects in the boundary $6d$ $\mathcal{N}= (0,2)$ theories, in terms of anomalies associated to the 2d surfaces that support them and from the bootstrap point of view. Whereas, \cite{FGT:2015} discusses the defects of the both types and shows the equivalence between them in the IR regime, in the context of $6d$ $\mathcal{N}= (0,2)$ theories where the M5 worldvolumes wrap Reimann surfaces of non-trivial topology $C$ $\subset$ $T^{\star}C$ inside a hyper-K\"ahler manifold with the special holonomy group $SU(2)$. The M-theoretic construction of such type was first discussed in \cite{Witten:97} which was based on the eleven-dimensional uplift of type IIA models of Hanany and Witten in \cite{HananyWitten:96}.
		
		 Wilson surface operators which are codimension-4 in the $6d$ $(0,2)$ tensor multiplet theory were also analysed in \cite{Lunin:2007, Chen:2007, MoriYamaguchi:2014} using the probe M5 branes in the $AdS_7 \times S^4$ geometry with the 3-form fluxes turned on along a $S^3$ sphere wrapped by their worldvolume. These Wilson surfaces become the Wilson line operators in the 5d $SU(N)$ SYM theory upon dimensional compactification along the circular direction. The probe M5 branes in these analysis had the worldvolume topology of $AdS_3 \times S^3$ and have the correspondence with the Wilson surfaces associated with the higher rank representations of the gauge theory in the large $N$ limit. 
		
		In our analysis, the discussion of probe M5 branes will be useful to shed more light on some features of the codimension-2 defects in the $6d$ $(0,2)$ theory. The probe M5 branes we will consider will have non-compact worldvolume, which extend to the boundary of the $AdS$ region and end in a four dimensional submanifold. Our goal in this manuscript is to derive the general M5 brane solutions that preserve the least amount of supersymmetry. And further, to give some general characterization equations that govern the behaviour, near the boundary of the $AdS$, of those BPS M5 solutions which wrap a circle direction in the $S^4$ part of the 11-dimensional geometry. 
		
		In the second section, we will begin by considering several classical solutions of $\frac 12$-BPS M5 probe branes in $AdS_7 \times S^4$ background expressed in the global coordinates. The equations that define the worldvolume of these probe M5s are analogous to the ones used for probe D3 brane solutions \cite{Drukker:2008wr, AGS:2020} which were used to study the BPS monodromy defects in $\mathcal{N}=4$ SYM theory holographically. These are noncompact brane solutions with world-volume toplogy of $AdS_5 \times S^1$ ending on the boundary in $\mathbb{R} \times S^3$ submanifold. They are one of the new findings that appear in this manuscript. The intersection of these M5 brane probes with the boundary should essentially be the half-BPS codimension-2 defects in the $6d$ $(0,2)$ theory on $\mathbb{R} \times S^5$. We perform a $\kappa$-symmetry analysis to find the projections on the bulk Killing spinor for the various $\frac12$-BPS probes. In section 3, we use the common supersymmetry of all the half-BPS probe solutions to derive the general BPS equations whose solutions preserve the least amount of supersymmetry from the ambient 11 dimensional geometry. A common single supersymmetry also comes from those half-BPS solutions which have compact world-volumes and are more commonly referred as the dual-giant gravitons in the literature \cite{Grisaru:2000zn, McGreevy:2000cw}. The worldvolume of the general probe branes are described by zeros of the holomorphic functions that satisfy a scaling condition. The general expression that we derive here give the worldvolumes of the least supersymmetric probe M5 branes in the $AdS_7 \times S^4$ geometry which include the dual-giant graviton configurations as well as the holographic dual of codimension-2 defects in the boundary 6d gauge theory. In section 4, we look at those M5 brane solutions that wrap a circle in the $S^4$ part of the geometry, they are in general $\frac1{16}$-BPS configurations. We do the analysis near the boundary of $AdS_7$, and we show how the zero locus of the holomorphic function should coincide with the location of codimension-2 defect in the $\frac1{16}$-BPS sector of the $6d$ $(0,2)$ theory. Moreover, we also propose that the $\frac1{16}$-BPS defects have a general characterisation which describes them as the intersection of the zeros of a holomorphic function $F(z_1, z_2, z_3) = 0$ with the 5-sphere defined by $|z_1|^2 + |z_2|^2 + |z_3|^2 = 1$. This analysis of the general $\frac1{16}$-BPS probe M5 brane near the boundary of $AdS$ also suggests us the kind of singular behavior the bosonic fields in the 6d theory may acquire near the location of the defect.
	\section{M5 branes in $AdS_7 \times S^4$ geometry}
		
		\noindent
		We consider the following metric of the eleven-dimensional $AdS_7 \times S^4$ geometry in global coordinates system
		\begin{align}
		\label{gddbulk}
		ds^2_{AdS} = - \left(1 + \frac{r^2}{4l^2}  \right) dt^2 + \frac{dr^2}{\left(1 + \frac{r^2}{4l^2} \right) } + r^2 d\Omega_5
		\end{align}
		with $d\Omega_5 = d\alpha^2 + \cos^2 \alpha \, d\phi_1^2  + \sin^2 \alpha \left( d\beta^2 + \cos^2 \beta \, d\phi_2^2 + \sin^2 \beta \, d\phi_3^2 \right) $
		\begin{align}
		\label{S4metric}
		ds^2_{S^4} =  l^2 \left( d\theta^2 +  \sin^2 \theta ( d\chi^2 + \cos^2 \chi \, d\xi_1^2  + \sin^2 \chi \, d\xi_2^2 ) \right)
		\end{align}
		%
		%
		%
		There is the 4-form flux $F^{(4)}$ through $S^4$ obtained from the potential
		\begin{align}
		A^{(3)} = \frac{l^3}2 \cos \theta \left( \cos 2 \theta \,-\, 5 \right)\, d\Omega_3
		\end{align}
		so that the field strength is $F^{(4)} = 3 \, l^3 \, d\Omega_4$. \\
		
		The global $AdS_7$ coordinates above can be written in terms of the following complex coordinates in $\mathbb{C}^{1,3}$
		\begin{align}
		\label{complexAdS7}
		\Phi_0 = l \cosh \rho  \, e^{i \phi_0} \quad \Phi_1 = l \sinh \rho \cos \alpha \, e^{i \phi_1} \quad  \Phi_2 = l \sinh \rho \sin \alpha \cos \beta \, e^{i \phi_2} \quad  \Phi_3 = l \sinh \rho \sin \alpha \sin \beta \, e^{i \phi_3}
		\end{align}
		which define the $AdS_7$ part as the following locus in $\mathbb{C}^{1,3}$
		\begin{align}
		- | \Phi_0 |^2 + | \Phi_1 |^2  + | \Phi_2 |^2 + | \Phi_3 |^2  = -  l^2
		\end{align}
		For the $S^3$ $\subset$ $S^4$ we define the complex cooordinates describing it embedded in $\mathbb{C}^2$ space
		\begin{align}
		Z_1 = \cos \chi \, e^{i \xi_1}  \quad \qquad \quad  Z_2 = \sin \chi \, e^{i \xi_2} \,.
		\end{align}
		\subsubsection*{Killing spinor of the background}
		We choose the frame vielbein such that it becomes manifest that the $AdS_7$ part ( respectively $S^3 \subset S^4$ part) can be written as a $U(1)$ Hopf fibration over a K\"ahler manifold $\widetilde{\mathbb{CP}}^3$ (respectively $\mathbb{CP}^1$). Here $\widetilde{\mathbb{CP}}^3$ is the hyberbolic version of the complex projective space and it is defined as the set of rays in the complex space $\mathbb{C}^{1,3}$( in place of the $\mathbb{C}^4$ space). The frame vielbein that we use are the following
		\begin{align}\label{adsframe}
		e^0 &= 2 l \left( \cosh^2 \rho \, d\phi_0 - \sinh^2 \rho \left( \cos^2 \alpha \, d\phi_1 + \sin^2 \alpha \cos^2 \beta \, d \phi_2 + \sin^2 \alpha \sin^2 \beta \, d \phi_3 \right)\right) \cr
		e^1 &= 2 l \, d \rho , \quad e^2 = 2 l  \sinh \rho \, d\alpha , \quad e^3 = 2 l \sinh \rho \sin \alpha \, d\beta \cr 
		e^4 &= 2 l \cosh \rho \sinh \rho \left( \cos^2 \alpha \, d\phi_{01} + \sin^2 \alpha \cos^2 \beta \, d\phi_{02} + \sin^2 \alpha \sin^2 \beta \, d\phi_{03} \right)   \cr
		e^5 &= 2 l \sinh \rho \cos \alpha \sin \alpha \left( \cos^2 \beta \, d\phi_{02} + \sin^2 \beta \, d\phi_{03} - d\phi_{01} \right) \cr
		e^6 &= 2 l \sinh \rho \sin \alpha \cos \beta \sin \beta \left( d\phi_{03} - d\phi_{02} \right)
		\end{align}
		%
		where $r = 2 l \sinh \rho$, $\phi_0 = \frac t{2l}$, and
		\begin{align}\label{S4frame}
		e^7 &= l \, d\theta , \quad e^8 = l\,\sin \theta  d\chi  , \quad e^9 = l \,\sin \theta \cos \chi \sin \chi \left(  d\xi_1 \, -  \, d\xi_2 \right) \,, \cr 
		e^{\underline{10}} &= l\,\sin \theta \left( \cos^2 \chi \, d\xi_1 + \sin^2 \chi \, d\xi_2 \right) \,.
		\end{align}
		\noindent
		The Killing spinor has to satisfy
		\begin{align}
		D_M \epsilon - \frac1{288} \left( \Gamma_M \Gamma^{PQRS} \, - \,  8 \delta_M^P \Gamma^{QRS} \right) F^{(4)}_{PQRS} \, \epsilon = 0 \,.
		\end{align}
		When $M$ is the index of one of the coordinates from $AdS_7$ the killing spinor equation reads
		\begin{align}
		D_{\mu} \epsilon - \frac1{4l} \Gamma_{\mu} \gamma \epsilon = 0 \,,
		\end{align}
		where $\gamma$ denotes the four-product $\Gamma_{789\underline{10}}$. When $M$ is the index of one of the coordinates from $S^4$ the equation reads
		\begin{align}
		D_m \epsilon + \frac1{2l} \Gamma_m \gamma \epsilon = 0 \,.
		\end{align}
		The solution for the Killing spinor with the above frame vielbein is given by :
		\begin{align}
		\label{adskss2}
		\epsilon &=  e^{\frac 12 ( \Gamma_{04} + \Gamma_1 \gamma ) \rho} e^{\frac 12 \left( \Gamma_{12} + \Gamma_{45} \right) \alpha} e^{\frac 12 \left( \Gamma_{23} + \Gamma_{56} \right) \beta}  e^{\frac 12 \Gamma_0 \gamma  \phi_0} e^{ - \frac 12 \Gamma_{14} \phi_1 } e^{ - \frac 12 \Gamma_{25} \phi_2} e^{ - \frac 12 \Gamma_{36} \phi_3} \cr
		& \hspace{1cm} \times e^{\frac 12 \gamma \Gamma_7 \theta } e^{ \frac 12 \left( \Gamma_{78} + \Gamma_{9\underline{10}} \right) \, \chi} e^{ \frac 12  \Gamma_{7\underline{10}} \xi_1 } e^{- \frac 12 \Gamma_{89} \xi_2} \epsilon_0 \equiv M \epsilon_0
		\end{align}
		The eleven-dimensional $\Gamma$-matrices satisfy the identity: $\Gamma_{01234\ldots\underline{10}} = 1$.  
		\vspace{.3cm}
		
	\subsection{Half-BPS M5 brane embeddings}\label{halfBPS}
We have several half-BPS solutions obeying the Euler-Lagrange equations of motion. We will divide them into the following two kind
\begin{enumerate}
	\item $\zeta_1 \Phi_i  =  l \sinh \rho_0 \, e^{i \xi_{const}}$
	\item $\zeta_2 \Phi_i  =  l \sinh \rho_0 \, e^{i \xi_{const}}$
\end{enumerate}
Here we define the variables $\zeta_i$ which are related to the complex coordinates that describe $S^3$: 
\begin{align}
	\label{zetavsZ}
   ( \zeta_1 )^2 = Z_1 \quad \quad \text{and} \quad \quad  ( \zeta_2 )^2 = Z_2 \,.
\end{align}

\noindent
There are four solutions that can be considered in each kind. And the solutions in the two kinds are related by an appropriate SU(2) rotation on the variables $Z_i$. For all the solutions that we consider in this section, we choose the static gauge in which the world-volume coordinates are identified as follow: $$ \tau = \phi_0 ,\,\, \sigma_1 = \alpha,\,\, \sigma_2 = \beta,\,\, \sigma_3 = \phi_1,\,\, \sigma_4 = \phi_2 \,\,\,\, \text{and} \,\,\,\, \sigma_5 = \phi_3 \,.$$

The Lagrangian density associated with all the probe M5 brane solutions that we discuss here is given by the following expression
\begin{align}
		\mathcal{L} ~= \,  T_5 \, \sqrt{-h} \, + \, T_5 \, P[ C^{(6)} ]
\end{align}
where $h$ is the determinant of the induced metric on the worldvolume and $P[\cdot]$ refers to
the pullback of a spacetime differential form onto the worldvolume. $C^{(6)}$ is the six-form gauge potential whose field strength $F^{(7)}$ is $\star$dual to the 4-form flux $F^{(4)}$ through $S^4$. We choose the gauge such that the six-form potential is given by
\begin{align}
		C^{(6)} = - \left( 2 l\right)^6  \sinh^6 \rho  \cos \alpha  \sin^3 \alpha  \cos \beta  \sin \beta \, d\alpha \wedge d\beta \wedge d\phi_0 \wedge \ldots \wedge d\phi_3 ~.
\end{align}
There are no world volume fluxes due to the gauge fields, and their value is zero throughout the analysis. \\

Our strategy in the remaining of this section is to analyze the half-BPS solutions (which are of the simpler form) and discuss the supersymmetry preserved by each in terms of the projection condition on the constant spinor $\epsilon_0$ in the killing spinor in \eqref{adskss2}. The probe M5 solutions of the non-compact world-volumes are new here and have never been presented before in the background geometry of global $AdS_7 \times S^4$. These solutions are the holographic dual of the half-BPS codimension-2 defects in the 6d boundary dual theory. In the next section, we make use of the crucial knowledge of the common supersymmetry preserved among all the $8$ solutions here to determine much more general BPS solutions preserving just a single supersymmetry. 
\vspace{.3cm}
\subsection*{Solutions of the $\textbf{I}^{st}$ kind}
First we consider the solutions in which the M5 brane world-volumes wrap a maximal circle on the $S^4$ part of the 11-dimensional geometry, parametrized by the coordinate $\xi_1$. 

\vspace{.6cm}
\subsubsection{ Solution: $\zeta_1 \Phi_0  = l \sinh \rho_0 \, e^{i \, \xi^{(0)}}$}
\label{halfBPSM5no0}
This solution describes the world-volume of a dual-giant graviton of \cite{McGreevy:2000cw} that wraps the $S^5$ sphere in the $AdS$ direction.
In terms of the real coordinates the defining equations are 
\begin{align}
\theta = \frac{\pi}{2} \qquad \chi = 0 \qquad \rho = \rho_0 (const.) \qquad 2 \phi_0 + \xi_1 =  \xi^{(0)}
\end{align}
The induced metric on the world-volume is
\begin{align}
ds^2 \bigg|_{M5} =& h_{\mu\nu} d\sigma^{\mu} d\sigma^{\nu} \cr
=&   4 l^2 \sinh^2 \rho_0 \left( -   d\phi_0^2 + d\alpha^2 + \cos^2 \alpha d\phi_1^2 + \sin^2 \alpha ( d\beta^2 + \cos^2 \beta d\phi_2^2 + \sin^2 \beta d\phi_3^2 )   \right) \,, \cr
\end{align}
it has a topology of $ \mathbb{R} \times S^5$. The square root of the determinant of the induced metric is equal to  $$\sqrt{-\det h} =  2^{5} l^6 \sinh^{6} \rho_0 \cos \alpha \sin^3 \alpha \sin 2\beta  \,.$$ 
%
%
The product of the six worldvolume $\gamma-$matrices is 
\begin{align}
\label{GammakPhi0}
\gamma_{\phi_0\alpha\beta\phi_1\phi_2\phi_3} = - 32\, l^6 \sinh^5 \rho_0 \cos \alpha \sin^3 \alpha \sin 2\beta \left( \cosh \rho_0 \, \Gamma_{023456} +  \cosh \rho_0 \, \Gamma_{23456\underline{10}} + \sinh \rho_0 \, \Gamma_{02356\underline{10}} \right)
\end{align}
The $\kappa$-symmetry equation that ensures the supersymmetry of the probe M5 is given by
\begin{align}
\Gamma_{\kappa} \epsilon = \epsilon
\end{align}
where the matrix $\Gamma_{\kappa}$ is equal to $$\Gamma_{\kappa} = \frac{ \gamma_{\phi_0\alpha\beta\phi_1\phi_2\phi_3}}{ \sqrt{- \det h} }\,.$$ \\
We subsititute for $\gamma_{\phi_0\alpha\beta\phi_1\phi_2\phi_3}$ from \eqref{GammakPhi0} in the kappa-symmetry equation. Then we commute each $\Gamma_{abcdef}$ product in \eqref{GammakPhi0} through the matrix factor $M$ in the killing spinor \eqref{adskss2}. 
%
%
The above solution satisfies the $\kappa$-symmetry constraint and preserves half of the eleven-dimensional supersymmetries if the following projection is considered on the constant spinor:
\begin{align}
\left( 1 - \Gamma_{089} \right) \epsilon_0 = 0 \,.
\end{align}
%
\subsubsection{ Solution: $\zeta_1 \Phi_1  = l \sinh \rho_0 \, e^{i \, \xi^{(0)}}$}
\label{halfBPSM5no1}
This solution describes the M5 branes with non-compact world-volumes that end in the boundary region of AdS into a four-dimensional submanifold. They are the holographic duals of codimension-2 defects in the 6d $(0,2)$ theory on $S^5 \times \mathbb{R}$ manifold. In terms of the real coordinates the embedding equations are
\begin{align}
\theta = \frac{\pi}2 \qquad \chi = 0 \qquad \sinh \rho \cos \alpha = \sinh \rho_0 \qquad 2 \phi_1 + \xi_1 = \xi^{(0)}
\end{align}
The metric induced on the world-volume of this classical solution is
\begin{align}
\label{inducedmetrichalfBPS1}
ds^2 \Big|_{M5} =& - 4 l^2 ( 1 + \sinh^2 \rho_0 \, \sec^2 \alpha ) d\phi_0^2  +  \frac{4 l^2 \sinh^2 \rho_0 \cosh^2 \rho_0 \sec^4 \alpha}{1 + \sinh^2 \rho_0 \, \sec^2 \alpha} d\alpha^2  + 4l^2  \sinh^2 \rho_0 \tan^2 \alpha \,  d\beta^2 \cr & + 4 l^2 \cosh^2 \rho_0 \, d\phi_1^2 + 4 l^2 \sinh^2 \rho_0 \tan^2 \alpha \cos^2 \beta \, \, d\phi_2^2 + 4 l^2 \sinh^2 \rho_0 \tan^2 \alpha \sin^2 \beta \, d\phi_3^2 \,.
\end{align}
\\
The metric is of topology $AdS_5 \times S^1$ and the scalar curvature is equal to $- \frac{5 \, \cosh^{-2} \rho_0}{l^2}$.
The square root of the determinant of the induced metric is equal to $$\sqrt{- \det h} =  2^{5} l^6 \sinh^4 \rho_0  \cosh^2 \rho_0  \sec^2 \alpha \tan^3 \alpha \sin 2\beta\,.$$ 
%
%
The product of the six worldvolume  $\gamma$-matrices is 
\begin{align}
\label{gammakappaHBPS1}
\gamma_{\phi_0\alpha\beta\phi_1\phi_2\phi_3}	=& - 16 \, l^6 \sinh^4 \rho_0 \sec^2 \alpha \tan^3 \alpha \sin 2\beta \times \bigg( ( 1 + 2 \sinh^2 \rho_0 + \cos 2 \alpha ) \, \Gamma_{02356\underline{10}}  \cr & \hspace{2.5cm} +  2 \sinh \rho_0 \cos \alpha  \sqrt{1+\sinh^2 \rho_0 \sec^2 \alpha} \, ( \Gamma_{023456} + \Gamma_{23456\underline{10}} ) \cr & \hspace{2.5cm} + 2 \sinh \rho_0 \sin \alpha \, ( \Gamma_{013456} + \Gamma_{13456\underline{10}} ) + 2 \sin^2 \alpha \, \Gamma_{01346\underline{10}} \cr
& \hspace{2.5cm} + \sin 2 \alpha \sqrt{1 + \sinh^2 \rho_0 \sec^2 \alpha} \, ( \Gamma_{01356\underline{10}} + \Gamma_{02346\underline{10}} )  \bigg)
\end{align}
%
We substitute this $\gamma_{\phi_0\alpha\beta\phi_1\phi_2\phi_3}$ in the $\kappa-$symmetry equation
%
%
%
and take each product of six $\Gamma$-matrices to the right through the exponential factors in the killing spinor denoted by $M$ in \eqref{adskss2}. After doing the $\Gamma$-matrix algebra, the l.h.s. in the $\kappa$-symmetry equation becomes 
\begin{align}
\Gamma_{\kappa} \cdot  M \cdot \, \epsilon_0 
=& \, M \cdot \Gamma_{0235689} \cdot \epsilon_0 \, + \, M \sin \beta \tan \alpha \tanh^2 \rho_0 \, e^{\Gamma_{14} \phi_1} e^{\Gamma_{36} \phi_3} \left( \Gamma_{0125689} + \Gamma_{02456789\underline{10}} \right) \cdot \epsilon_0  \cr
& - M \cos \beta \tan \alpha \tanh^2 \rho_0 \, e^{\Gamma_{14} \phi_1} e^{\Gamma_{25} \phi_2} \left( \Gamma_{0135689} + \Gamma_{03456789\underline{10}} \right) \cdot \epsilon_0 \cr
& - M \tanh \rho_0 \cosh^{-1} \rho_0 \sqrt{1 + \sinh^2 \rho_0 \sec^2 \alpha} \, e^{- \Gamma_0 \gamma \phi_0} e^{\Gamma_{14} \phi_1} \left( \Gamma_{0123567\underline{10}} + \Gamma_{023456} \right) \cdot \epsilon_0 
\end{align}
The $\kappa-$symmetry equation is satisfied and half of the eleven-dimensional supersymmetries are preserved by the M5 world-volume, when the following projection is applied on the constant spinor $\epsilon_0$
\begin{align}
\left( 1 - \Gamma_{147\underline{10}} \right) \epsilon_0 = 0 \,.
\end{align}
The remaining two solutions in this kind can be obtained by doing the appropriate $SU(3)$ transformations on the complex coordinate $\Phi_1$.

\vspace{.3cm}
\subsubsection{Solution: $\zeta_1 \Phi_2  =  l \sinh \rho_0 \, e^{i \, \xi^{(0)}}$}
\label{halfBPSM5no2}
This is an another example of non-compact worldvolume solution which end in the AdS boundary into a 4-dimensional submanifold. This will also be a holographic dual of a codimension-2 defect in the 6d $(0,2)$ theory. In terms of the real coordinates the embedding equations are
\begin{align}
\theta = \frac{\pi}2 \qquad \chi = 0 \qquad \sinh \rho \, \sin \alpha \cos \beta = \sinh \rho_0 \qquad 2 \phi_2 + \xi_1 = \xi^{(0)}
\end{align}
\noindent
%
%
The induced metric again is of topology $AdS_5 \times S^1$ and the scalar curvature of the above metric is $- \frac{5 \, \cosh^{-2} \rho_0}{l^2}$. 
With the determinant: $$\det h = - (2 l )^{12} \sinh^8 \rho_0 \cosh^4 \rho_0 \cot^2 \alpha \csc^4 \alpha \sec^8 \beta \tan^2 \beta.$$ 
The above M5 brane solution satisfies the $\kappa-$symmetry constraint  
\begin{align}
\Gamma_{\kappa} \epsilon =  \epsilon \,,
\end{align}
and half of the eleven-dimensional supersymmetries are preserved by the world-volume, when the following projection is applied on the constant spinor $\epsilon_0$
\begin{align}
\left( 1 - \Gamma_{257\underline{10}} \right) \epsilon_0 = 0 \,.
\end{align}
%
\subsubsection{Solution: $\zeta_1 \Phi_3  =  l \sinh \rho_0 \, e^{i \, \xi^{(0)}}$}
\label{halfBPSM5no3}
This is the fourth half-BPS solution in the first set of solutions. In terms of the real coordinates the embedding equations are
\begin{align}
\theta = \frac{\pi}2 \qquad \chi = 0 \qquad \sinh \rho \, \sin \alpha \sin \beta = \sinh \rho_0 \qquad 2 \phi_3 + \xi_1 = \xi^{(0)}
\end{align}
The above solution will also preserve half the supersymmetries if the following projections on $\epsilon_0$ are imposed
\begin{align}
( 1 - \Gamma_{367\underline{10}} ) \epsilon_0 = 0 \,.
\end{align}
\subsection*{Solutions of the $\textbf{II}^{nd}$ kind}
We now consider the solutions which wrap the maximal circle on $S^4$ parametrized by the coordinate $\xi_2$.

\vspace{.6cm}
\subsubsection{Solution: $\zeta_2 \Phi_0  =  l \sinh \rho_0 \, e^{i \, \xi^{(0)}}$}
This solution describes a dual-giant graviton with worldvolume wraping the $S^5$ sphere in the AdS directions. In terms of the real coordinates the embedding equations are
\begin{align}
\theta = \frac{\pi}2 \qquad \chi = \frac{\pi}2 \qquad \rho= constant \qquad 2 \phi_0 + \xi_2 = \xi^{(0)}
\end{align}
The above solution will also preserve half of the supersymmetries if the following projections on $\epsilon_0$ are imposed
\begin{align}
( 1 + \Gamma_{07\underline{10}} ) \epsilon_0 = 0 \,.
\end{align}
\subsubsection{Solution: $\zeta_2 \Phi_1  =  l \sinh \rho_0 \, e^{i \, \xi^{(0)}}$}
This solution describes a non-compact worldvolume brane with holographic duality to a codimension-2 defect in the 6d theory. In terms of the real coordinates the embedding equations are
\begin{align}
\theta = \frac{\pi}2 \qquad \chi = \frac{\pi}2 \qquad \sinh \rho \cos \alpha = \sinh \rho_0 \qquad 2 \phi_1 + \xi_2 = \xi^{(0)}
\end{align}
The above solution will also preserve half the supersymmetries if the following projections on $\epsilon_0$ are imposed
\begin{align}
( 1 + \Gamma_{1489} ) \epsilon_0 = 0 \,.
\end{align}
Similar to the first set of solutions there are two more solutions that can be obtained by doing the appropriate $SU(3)$ rotations on the coordinate $\Phi_1$.
\subsubsection{Solution: $\zeta_2 \Phi_2  =  l \sinh \rho_0 \, e^{i \, \xi^{(0)}}$}
In terms of the real coordinates the embedding equations are
\begin{align}
\theta = \frac{\pi}2 \qquad \chi = \frac{\pi}2 \qquad \sinh \rho \sin \alpha \cos \beta = \sinh \rho_0 \qquad 2 \phi_2 + \xi_2 = \xi^{(0)}
\end{align}
The above solution will also preserve half the supersymmetries if the following projections on $\epsilon_0$ are imposed
\begin{align}
( 1 + \Gamma_{2589} ) \epsilon_0 = 0 \,.
\end{align}
\subsubsection{Solution: $\zeta_2 \Phi_3  =  l \sinh \rho_0 \, e^{i \, \xi^{(0)}}$}
In terms of the real coordinates the embedding equations are
\begin{align}
\theta = \frac{\pi}2 \qquad \chi = \frac{\pi}2 \qquad \sinh \rho \sin \alpha \sin \beta = \sinh \rho_0 \qquad 2 \phi_3 + \xi_2 = \xi^{(0)}
\end{align}
The above solution will also preserve half the supersymmetries if the following projections on $\epsilon_0$ are imposed
\begin{align}
( 1 + \Gamma_{3689} ) \epsilon_0 = 0 \,.
\end{align}
\subsection{Some $\frac14$-BPS solutions}\label{quarterBPS}
From the analysis we have seen so far we can consider to combine two of the solutions from above. This time the parametrization of the 1-dimensional curve that is wrapped on the $S^4$ can be understood by using both the coordinates: $\xi_1$ and $\xi_2$.  \\

\noindent
The complete ansatz for an M5-brane world-volume made from the set of solutions of the Euler-Lagrange equation is 
\begin{align}
	\theta = \frac{\pi}2 \qquad  \sinh \rho \cos \alpha = \sinh \rho_0 \qquad 2 \phi_1 + \xi_1 = 0 \quad 2 \phi_1 + \xi_2 = 0 \,.
\end{align}
Along with the $S^4$ coordinate '$\chi$' fixed to an arbitrary value(other than $0$ or $\frac{\pi}2$). In terms of the special complex variable we have defined  in \eqref{zetavsZ} this ansatz takes the form: $\zeta_1 \Phi_1 = c_1$ and $\zeta_2 \Phi_1 = c_2$ with $c_1$, $c_2$ some arbitrary constants. This choice of ansatz indicates that the curve that is wrapped on the $S^4$ by the probe M5 solution is parametrized by the Hopf fibre direction coordinate in which the frame component $e^9$ vanishes and $e^{\underline{10}}$ becomes $l\, \left( \cos^2 \chi \, d\xi_1 + \sin^2 \chi \, d\xi_2 \right) = l \, d\xi_1 \,( \text{or} \,\,  l \, d\xi_2 )$.\\

\noindent
The induced metric on the world-volume remains the same as in \eqref{inducedmetrichalfBPS1}
 and the product of the six $\gamma$ matrices to be used in the kappa-symmetry equation is also the same as in \eqref{gammakappaHBPS1}. After pushing through all the six-product $\Gamma_{ab\ldots f}$ matrices in \eqref{gammakappaHBPS1} through the matrix factor $M$ in the killing spinor in \eqref{adskss2}, the kappa-symmetry constraint: $\Gamma_{\kappa} \epsilon = \epsilon$ can be made to be satisfied as in the previous 8 half-BPS solutions we have seen so far. But this time we need to impose two independent projection conditions in order to accomplish this. It means that the worldvolume now preserves the quarter of the 11-dimensional spacetime supersymmetries and the answer in terms of the projection conditions is the following \\
\begin{align}
( 1 - \Gamma_{147\underline{10}} ) \epsilon_0  \, = \, 0  \qquad \qquad ( 1 - \Gamma_{789\underline{10}} ) \epsilon_0  \, = \, 0 \,.
\end{align}
\\
\noindent
Likewise, there are other solutions which can be considered with the ansatz $\zeta_1 \Phi_i = c_1$ and $\zeta_2 \Phi_i = c_2$ with $i = 0, 2,$ and $3$. Each of these world-volume solutions are BPS and preserve a quarter of the 11-d supersymmetries, with the second projection condition being the same and the first condition being altered according to the illustrations given in previous subsection in \ref{halfBPSM5no0}, \ref{halfBPSM5no2} and \ref{halfBPSM5no3}, respectively. The world volume metric and the the six-product $\gamma_{\tau \sigma_1 \ldots \sigma_5}$ of the world-volume $\gamma$-matrices remains unchanged as given for the respective cases in \ref{halfBPSM5no0}, \ref{halfBPSM5no2} and \ref{halfBPSM5no3}. For convenience, we write the projection conditions for the each individual $\frac 14$-BPS solutions separately \\
\begin{align}
\zeta_1 \Phi_0 = c_1 \,\,;\,\, \zeta_2 \Phi_0 = c_2\,\quad:& \qquad ( 1 - \Gamma_{089} ) \epsilon_0  \, = \, 0  \quad \quad ( 1 - \Gamma_{789\underline{10}} ) \epsilon_0  \, = \, 0  \cr
\zeta_1 \Phi_2 =  c_1 \,\,;\,\, \zeta_2 \Phi_2 = c_2\,\quad:& \qquad ( 1 - \Gamma_{257\underline{10}} ) \epsilon_0  \, = \, 0  \quad \quad ( 1 - \Gamma_{789\underline{10}} ) \epsilon_0  \, = \, 0 \cr
\zeta_1 \Phi_3 = c_1\,\,;\,\, \zeta_2 \Phi_3 = c_2\,\quad:& \qquad ( 1 - \Gamma_{367\underline{10}} ) \epsilon_0  \, = \, 0  \quad \quad ( 1 - \Gamma_{789\underline{10}} ) \epsilon_0  \, = \, 0 \,.
\end{align}
\section{The General BPS solutions}
In this section, we consider the projections that preserve the common set of supersymmetries amongst all the eight half-BPS probe M5 solutions that we have seen so far, by keeping both $\zeta_1$ and $\zeta_2$ non-zero. It is given by the following set of projections on the constant spinor: \\
\begin{align}
\label{1by32BPSprojections}
\Gamma_{14} \epsilon_0 = \Gamma_{25} \epsilon_0 = \Gamma_{36} \epsilon_0 = i \, \epsilon_0~\, \quad \text{and} \qquad   \Gamma_0 \epsilon_0  =  \Gamma_{7\underline{10}} \epsilon_0 = - \Gamma_{89} \epsilon_0 =   - i \, \epsilon_0
\end{align}
\\
These projections preserve just one out of the thirty two supersymmetries of the bulk background.
After taking into account the action of these projection conditions on $\epsilon_0$, the killing spinor simplifies to 
\begin{align}
\label{adsKS1by32simplified}
\epsilon = e^{  - \frac i2 \left( \phi_0 + \phi_1 + \phi_2 + \phi_3 + \xi_1 + \, \xi_2 \right) } \, \frac{ 1 - \Gamma_7 }{\sqrt{2}} \, \epsilon_0 \,.
\end{align}
In addition to this simplified form of the killing spinor, the action of any six-product $\Gamma$-matrix: $ \Gamma_{abcde\underline{10}}$( with $a,b,c,d,e \, \ne \, 7,\underline{10}$ ) on $\epsilon$ effectively becomes
\begin{align}
\Gamma_{abcde\underline{10}} \, \epsilon = - i \, e^{  - \frac i2 \left( \phi_0 + \phi_1 + \phi_2 + \phi_3 + \xi_1 + \, \xi_2 \right) } \, \frac{ 1 - \Gamma_7 }{\sqrt{2}} \, \Gamma_{abcde} \, \epsilon_0 \,.
\end{align}

\noindent
This relation is very crucial in deriving the constraint equations that we are after. We are now interested in finding the world volume solution for the general case, the expressions which describe the M5 branes that preserve atleast a single supersymmetry obtained from the 5 independent projection conditions that we have written in \eqref{1by32BPSprojections}. We follow the method used in the reference \cite{Ashok:2008fa} and introduce the complex 1-forms \\
\begin{align}
\label{complex-1form1}
\mathbf{E}^1 = \mathfrak{e}^1  - i \, \mathfrak{e}^4 \qquad & \qquad \mathbf{E}^2 = \mathfrak{e}^2  - i \, \mathfrak{e}^5 \cr \mathbf{E}^3 = \mathfrak{e}^3  - i \, \mathfrak{e}^6 \qquad & \qquad  \mathbf{E}^8 = \mathfrak{e}^8  - i \, \mathfrak{e}^{9} \,,
\end{align}
\\
where $\mathfrak{e}^a_i = e^a_{\mu} \partial_i X^{\mu}$ is the pullback of the spacetime frame $e^a_{\mu}$ onto the M5 world-volume. \\

\noindent
We substitute the simplified Killing spinor in \eqref{adsKS1by32simplified} in the $\kappa$-symmetry equation \\
\begin{align}
\label{kappasym2}
\gamma_{\tau \sigma_1 \sigma_2 \sigma_3 \sigma_4 \sigma_5} \epsilon = \pm  \sqrt{ - \det h } \, \epsilon \,,
\end{align}
\\
and use the projection conditions to reduce the LHS into a linear combination of independent structures of the form $\Gamma_{a_1 a_2 \ldots} \epsilon_0$. The coefficient of each such structure is set to zero except the constant one, which can be proved to equal to the RHS. In this section we are looking for solutions with the $S^4$ coordinate $\theta$ fixed to $\frac {\pi}2$, therefore, the 1-form $\mathfrak{e}^7$ is 0 and will never appear in the relations we obtain.
\\

On account of the projection conditions in \eqref{1by32BPSprojections} we find that there are various six-form differential constraints that need to be satisfied to obtain the solution we seek. To give a clear exposition of the steps involved in arriving at these BPS differential-form constraints we divide the ${10 \choose 6}$ number of terms in the l.h.s. of \eqref{kappasym2} into three different groups. In the first group, we only look at terms which either have $\mathfrak{e}^0$ or $\mathfrak{e}^{\underline{10}}$. The second group have terms with both $\mathfrak{e}^0$ and $\mathfrak{e}^{\underline{10}}$ present. And in the third group terms neither of the two are present. \\

\noindent
From the terms in the first group we get the following set of 6-form constraints
\begin{align}
\label{sixformBPSconstraint1}
	&\left( \mathfrak{e}^0 + \mathfrak{e}^{\underline{10}} \right) \wedge \overline{\mathbf{E}^a} \,  \overline{\mathbf{E}^b} \, \overline{\mathbf{E}^c} \wedge \left( \tilde{\omega} \, + \, \omega \right) = 0    \cr
	&\left( \mathfrak{e}^0 + \mathfrak{e}^{\underline{10}} \right) \wedge \overline{\mathbf{E}^a} \wedge  \left( \tilde{\omega} \, + \, \omega \right) \wedge  \left( \tilde{\omega} \, + \, \omega \right) = 0 \quad \,\, \qquad    \text{for} \,\, a,b,c = 1,2,3,8
\end{align}
with the definitions: $E^0 = \mathfrak{e}^{0} + \mathfrak{e}^{\underline{10}}$ and $\overline{E^0} = \mathfrak{e}^{0} - \mathfrak{e}^{\underline{10}}$, we re-write the above sets of constraints as \\
\begin{align}
\label{sixformBPSconstraint1pt2}
&E^0 \wedge \overline{\mathbf{E}^a} \,  \overline{\mathbf{E}^b} \, \overline{\mathbf{E}^c} \wedge \left( \tilde{\omega} \, + \, \omega \right) = 0    \cr
&E^0 \wedge \overline{\mathbf{E}^a} \wedge  \left( \tilde{\omega} \, + \, \omega \right) \wedge  \left( \tilde{\omega} \, + \, \omega \right) = 0 \quad \,\, \qquad    \text{for} \,\, a,b,c = 1,2,3,8
\end{align}
\\
Here we have also defined the following real 2-forms:
\begin{align}
		\tilde{\omega} = \mathfrak{e}^{14} + \mathfrak{e}^{25} + \mathfrak{e}^{36} = \frac i2 \left( \overline{\mathbf{E}^1} \, \mathbf{E}^1 \,+\, \overline{\mathbf{E}^2} \, \mathbf{E}^2 \,+\, \overline{\mathbf{E}^3} \, \mathbf{E}^3 \right) \equiv \omega_{\widetilde{\mathbb{CP}}^3}
\end{align}
\begin{align}
		\omega = \mathfrak{e}^{89} = \frac i2 \overline{\mathbf{E}^8} \, \mathbf{E}^8  \equiv \omega_{\mathbb{CP}^1} \,.
\end{align}
These 2-forms are the pull-backs of certain K\"ahler forms onto the worldvolume of the brane. These K\"ahler forms are of the respective base manifolds $\mathbb{CP}^1$ and $\widetilde{\mathbb{CP}}^3$, when the $S^3$ $\subset S^4 $ and $AdS_7$ are written as Hopf-fibrations. \\

The terms with a factor $\mathfrak{e}^{0\underline{10}}$ give the constraints
\begin{align}
		\label{sixformBPSconstraint2}
		&\mathfrak{e}^0 \wedge \mathfrak{e}^{\underline{10}} \wedge \overline{\mathbf{E}^a} \, \overline{\mathbf{E}^b} \wedge \left( \tilde{\omega} \, + \, \omega \right)  =  0 \cr
		& \mathfrak{e}^0 \wedge \mathfrak{e}^{\underline{10}} \wedge   \overline{\mathbf{E}^1} \,  \overline{\mathbf{E}^2}\,  \overline{\mathbf{E}^3}\,  \overline{\mathbf{E}^8} = 0 \,.
\end{align}
The BPS differential 6-form constraints from the remaining set of terms are
\begin{align}
		\label{sixformBPSconstraint3}
		 \overline{\mathbf{E}^a} \,  \overline{\mathbf{E}^b}\wedge (\omega \, + \, \tilde{\omega}) \wedge  (\omega \, + \, \tilde{\omega}) = 0 \, \qquad \,\,\,\, \quad \qquad \text{for} \,\, a,b = 1,2,3,8\,.
\end{align}
And finally, coefficients of the terms in \eqref{kappasym2} with all the six $\Gamma$ matrices projected out, suggest that they are equal to
\begin{align}
		   \left( \mathfrak{e}^{0} \wedge \mathfrak{e}^{\underline{10}} \, - \, i \, \frac{\omega \, + \, \tilde{\omega}}3 \right) \wedge \frac{ (\omega \, + \, \tilde{\omega}) \wedge \left( \tilde{\omega} \, + \, \omega \right)}2  \, = \,\,   \,  \sqrt{ - \det h }  \quad =   \, \text{dvol}_6 \,.
\end{align}
Where $\text{dvol}_6$ is the volume element on the world volume of the M5-brane. \\

\noindent
To describe the embedding of an M5 brane in 11-dimensional spacetime we need 5 real conditions. Fixing $\theta = \frac {\pi}2$ is one of them. To determine the other four we consider two holomorphic functions of spacetime coordinates whose zeros will give us the desired embedding solution
\begin{align}
F^{(I)} \left( \rho, \alpha, \beta, \phi_0, \phi_1, \phi_2, \phi_3, \chi, \xi_1 , \xi_2\right) = 0 \qquad \text{for} \,\, I=1,2
\end{align}
This leads to the differential constraints
\begin{align}
P \left[ F^{(I)}_{\rho} \, d\rho + F^{(I)}_{\alpha} \, d\alpha  + F^{(I)}_{\beta}  \, d\beta + \sum_{i=0}^3 F^{(I)}_{\phi_i} \, d\phi_i + \sum_{x^j=\chi, \xi_1, \xi_2}F^{(I)}_{x^j} \, dx^j \right] = 0
\end{align}
where $P$ denotes pullback onto the world-volume. We rewrite this in terms of the complex one-forms defined in \eqref{complex-1form1} using the frame vielbeins in \eqref{adsframe} and \eqref{S4frame}
\begin{align} \label{onedifferentialconstraint}
& \mathbf{E}^1 \left( F^{(I)}_{\rho} - i  \sum_{i=1,2,3} F^{(I)}_{\phi_i} \, \coth \rho - i \, F^{(I)}_{\phi_0} \tanh \rho \right)  
+  \overline{\mathbf{E}^1} \left( F^{(I)}_{\rho} + i  \sum_{i=1,2,3} F^{(I)}_{\phi_i} \, \coth \rho + i \, F^{(I)}_{\phi_0} \tanh \rho \right) \cr
&\sinh^{-1} \rho \left[ \mathbf{E}^2 \left(  F^{(I)}_{\alpha} - i \sum_{i=2,3} F^{(I)}_{\phi_i} \cot \alpha + i \, F^{(I)}_{\phi_1} \tan \alpha \right) + \overline{\mathbf{E}^2} \left(  F^{(I)}_{\alpha}  + i \sum_{i=2,3} F^{(I)}_{\phi_i} \cot \alpha - i \, F^{(I)}_{\phi_1} \tan \alpha \right) \right] \cr
& \, \csc \alpha  \sinh^{-1} \rho \bigg[ \mathbf{E}^3  \left( F^{(I)}_{\beta} - i \, F^{(I)}_{\phi_3} \cot \beta + i \, F^{(I)}_{\phi_2} \tan \beta \right) + \overline{\mathbf{E}^3} \left( F^{(I)}_{\beta} + i \, F^{(I)}_{\phi_3} \cot \beta - i \, F^{(I)}_{\phi_2} \tan \beta \right) \bigg] \cr 
&  \hspace{1cm} 2 \,  \mathbf{E}^8 \, \left( F^{(I)}_{\chi} - i \, F^{(I)}_{\xi_2} \cot \chi + i \, F^{(I)}_{\xi_1} \tan \chi \right)  + 
 2  \, \overline{\mathbf{E}^8} \, \left( F^{(I)}_{\chi} + i \, F^{(I)}_{\xi_2} \cot \chi - i \, F^{(I)}_{\xi_1} \tan \chi \right) \cr
 & \hspace{4cm}  +   \, E^0 \left(   \sum_{i=0}^3 F^{(I)}_{\phi_i} + 2 \sum_{i=1,2} F^{(I)}_{\xi_i} \right)  +  \, \overline{E^0} \left(  \sum_{i=0}^3 F^{(I)}_{\phi_i} - 2 \sum_{i=1,2} F^{(I)}_{\xi_i} \right)   = 0
\end{align}
On account of the constraint obtained in \eqref{sixformBPSconstraint1pt2}, \eqref{sixformBPSconstraint2} and \eqref{sixformBPSconstraint3} it can be checked that the l.h.s. can be made to zero if it is multiplied by appropriate factors, when there are a few additional differential equations are satisfied, that can be read from above. We write those down below to determine the functional form of $F^{(I)}$
\begin{align}
F^{(I)}_{\rho} + i  \sum_{i=1,2,3} F^{(I)}_{\phi_i} \, \coth \rho + i \, F^{(I)}_{\phi_0} \tanh \rho &= 0 \cr
F^{(I)}_{\alpha}  + i \sum_{i=2,3} F^{(I)}_{\phi_i} \cot \alpha - i \, F^{(I)}_{\phi_1} \tan \alpha  &= 0 \cr
F^{(I)}_{\beta} + i \, F^{(I)}_{\phi_3} \cot \beta - i \, F^{(I)}_{\phi_2} \tan \beta &= 0 \cr
F^{(I)}_{\chi} + i \, F^{(I)}_{\xi_2} \cot \chi - i \, F^{(I)}_{\xi_1} \tan \chi  &= 0 \,.
\end{align}	
The functional conditions have holomorphic dependence on $AdS_7$ complex coordinates and the complex coordinates describing the $S^3$ $\subset$ $S^4$
\begin{align}
\label{generalsolutions1by32}
F^{(I)} ( \Phi_0 \,, \Phi_1\,, \Phi_2 \,, \Phi_3 \,, Z_1, Z_2) = 0 \,.
\end{align}
There is a fifth differential equation that can also be read from \eqref{onedifferentialconstraint}
\begin{align}
	\label{scalingcondition}
	\sum_{i=0}^3 F^{(I)}_{\phi_i} - 2 \sum_{i=1,2} F^{(I)}_{\xi_i} \,= \, 0 \,,
\end{align}
 for $I$ equal to $1$ and $2$. We refer to it as the scaling condition on the holomorphic functions $F^{(I)}$ for our general solutions. It can be checked that from this general solution all the half-BPS solutions that we have discussed in section \ref{halfBPS} and $\frac14$-BPS solutions in section \ref{quarterBPS} can be derived easily. For the half-BPS solutions, the holomorphic functional constraints become 
 $$\left( F^{(1)} \left( \Phi_i \, ,Z_1 \right) = 0 \,; \, F^{(2)} = Z_2 = 0 \right) \,\, \text{or} \,\, \left( F^{(1)} = Z_1 =  0\,;\,F^{(2)} \left( \Phi_i \,, Z_2 \right) = 0 \right).$$ 
 And the scaling condition on $F^{(a)}$ makes them equivalent to the constraints $$ \zeta_a \Phi_i  - c_a  = 0 \, ; \, Z_b = 0 ~,$$ for the respective cases. 
 For the $\frac14$-BPS solutions, the holomorphic functional constraints become $$ F^{(1)} \left( \Phi_i \,, Z_1 \right) = 0 \,;\, F^{(2)} \left( \Phi_i \,, Z_2 \right) = 0 \,.$$ The scaling condition on $F^{(I)}$ makes them equivalent to $$\zeta_1 \Phi_i  - c_1  = 0 \, ; \,  \zeta_2 \Phi_i - c_2  = 0~,$$ for the respective cases. \\
 
 Similar equations were obtained in \cite{Ashok:2008fa} where the investigations were done for general D3 brane solutions in the $AdS_5 \times S^5$ spacetime. In \cite{AGS:2020} it was shown that the holographic duals of codimension-2 defects in 4d $\mathcal{N}=4$ SYM belong to same class of BPS solutions that also admit giant gravitons and the dual-giant gravitons from \cite{Ashok:2008fa}. In \cite{Mikhailov:2001} general giant M2 brane solutions which have two non-zero spin charges along $S^4$ were discussed in $AdS_7 \times S^4$ and they were atleast $\frac 14$-BPS. These solutions were obtained when at any given time the spatial part of M2 worldvolume is given by the intersection of a holomorphic curve $C$ in $\mathbb{C}^2$ space with $S^4$ sphere. Where the $S^4$ was considered to be embedded in the $5d$ flat space $\mathbb{R}^5 = \mathbb{C}^2 \times \mathbb{R}$.  

 From the general expression \eqref{generalsolutions1by32} that we have derived, depending upon the functional form of $F^{(I)}$ we can have the dual-giant gravtion M5 branes or the holographic duals of the codimension-2 defects in the boundary gauge theory that preserve the least supersymmetry from the projections given in \eqref{1by32BPSprojections}.
 \section{Boundary profile of the BPS solutions}
 
 The particular probe M5 brane solutions we are interested in and which will be the holographic duals of codimension-2 defects in the 6d theory have the charges $(S_1, S_2, S_3, J)$. Here $S_i$ denotes the spin angular momentum charges in the $AdS$ directions and $J$ is due to the spin in the $S^3 \subset S^4$ directions. In terms of the general expressions we derived, these solutions are described by 
 \begin{align}
 		\label{1by16BPSsolutionform}
 		Z_2 = 0 \quad \text{and} \quad F\left( \zeta \Phi_0 , \zeta \Phi_1, \zeta \Phi_2, \zeta \Phi_3 \right) = 0 \,,
 \end{align}
 where $\zeta = \sqrt{Z_1} = e^{i \frac{\xi_1}2 }$. The scaling condition \eqref{scalingcondition} has allowed us to write this solution with $Z_2 = 0$ in this form. It also means that the configuration is invariant under the following change
 \begin{align}
 	\Phi_i \rightarrow \lambda \Phi_i  \quad \text{and} \quad \zeta \rightarrow \lambda^{-1} \zeta \,. 
 \end{align}
 %
 The goal in this section is to find what functional form $F$ leads to near the boundary of $AdS_7$, and this will tell us about the location of the singularity along with the singular profile the 6d bosonic fields may acquire near the defect. As we near the boundary the complex coordinates $\Phi_i$ $\in$ $\mathbb{C}^{1,3}$ chosen for $AdS_7$ embedding take the form
 \begin{align}
 	\Phi_0 = r \nu_0 \quad \Phi_1 = r \nu_1 \quad \Phi_2 = r \nu_2 \quad \Phi_3 = r \nu_3 \,
 \end{align}
 where $r = 2 l \, \sinh \rho$ and $\nu_0 = e^{i\phi_0}$, $\nu_1 = \cos \alpha e^{i \phi_1}$, $\nu_2 = \sin \alpha \cos \beta e^{i \phi_2}$, $\nu_3 = \sin \alpha \sin \beta e^{i \phi_3}$. And in this large r approximation these become the coordinates on a null cone: $ -|\Phi_0|^2 + |\Phi_1|^2 + |\Phi_2|^2 + |\Phi_3|^2 = 0$. The induced metric on this cone tells us that the boundary is in the conformal class of $\mathbb{R} \times S^5$ for arbitrary and lagre r
 \begin{align}
 		- |d\Phi_0|^2 + |d\Phi_1|^2 + |d\Phi_2|^2 &+ |d\Phi_3|^2 = \cr 
 		&r^2 \left( - d\phi_0^2 + d\alpha^2 + \cos^2 \alpha \, d\phi_1 + \sin^2 \alpha \left( d\beta^2 + \cos^2 \beta \, d\phi_2^2 + \sin^2 \beta \, d\phi_3^2 \right) \right)  \cr
 \end{align}
Our steps are going to be very similar to the steps that were done in \cite{AGS:2020} for the general $\frac 18-$BPS non-compact D3 worldvolume solutions with a single spin angular momentun charge turned on in the $S^5$ part. Here we need to find the locus of zeros of the function in \eqref{1by16BPSsolutionform} as we near the boundary. Near the boundary the function becomes $F\left( \zeta r \nu_0,\, \zeta r \nu_1, \, \zeta r \nu_2, \, \zeta r \nu_3  \right)$ with $\zeta = e^{i \frac{\xi_1}2}$. So the worldvolume of the M5 brane intersects the boundary at the zeros of the functions
 \begin{align}
 		F\left( \lambda \nu_0,\, \lambda \nu_1, \,\lambda \nu_2, \, \lambda \nu_3  \right) = 0 \,,
 \end{align}
 where $\lambda = r \, e^{i \frac{\xi_1}2}$ for arbitrary $\lambda \in \mathbb{C}^*$. In this near boundary limit where $\Phi_i \rightarrow r \, \nu_i$, the zero set remains invariant only if this function is homogeneous under scaling, which means $ F\left( \lambda \nu_0,\, \lambda \nu_1, \,\lambda \nu_2, \, \lambda \nu_3  \right) = \lambda^p F\left(  \nu_0,\, \nu_1, \, \nu_2, \,  \nu_3  \right)$. And a function with such a scaling property can be re-written as $\nu_0^p \, H\left( \nu_1/ \nu_0 , \, \nu_2/\nu_0 , \, \nu_3/\nu_0 \right)$ so therefore, the zeros that we are after at a fixed 'time' $\tau$(or $\phi_0$) is the same as the zeros of the holomorphic function $H \left( \Lambda_1, \Lambda_2, \Lambda_3  \right)$ where $\Lambda_i = \frac{\nu_i}{\nu_0}$ $\in$ $\mathbb{C}^{3}$ which intersects the unit 5-sphere $|\Lambda_1|^2 + |\Lambda_2|^2 + |\Lambda_3|^2 = 1$. The time evolution here is given by the scaling $(\Lambda_1, \Lambda_2, \Lambda_3)$ $\rightarrow$ $e^{-i\phi_0} (\Lambda_1, \Lambda_2, \Lambda_3)$.  
 
 We summarize our result that we have derived just now for our $\frac 1{16}$-BPS probe M5 solutions. If the worldvolume of the non-compact probe M5 brane are described by the zero locus of an arbitrary function $F\left( \zeta \Phi_0 , \zeta \Phi_1, \zeta \Phi_2, \zeta \Phi_3 \right)$, then we see that the worldvolume, as it approaches the boundary is four dimensional and at a given instant in time, it is given by the locus $\mathcal{S}$ which is obtained as the intersection of a holomorphic function in $\mathbb{C}^3$ with the 5-sphere
 \begin{align}
 		\label{codimension2location}
		H \left( \Lambda_1, \Lambda_2, \Lambda_3  \right) = 0   \quad \cap \quad  |\Lambda_1|^2 + |\Lambda_2|^2 + |\Lambda_3|^2 = 1 \,.
 \end{align}
  This is certainly a higher dimensional generalization of the result (refer \cite{Milnor} for definition for instance), that we have seen in \cite{AGS:2020} in terms of the algebraic link in $S^3$, obtained for the probe M5 brane in the $AdS_7 \times S^4$ background.
  
  We can also comment on the profile the bosonic fields in 6d $\mathcal{N}=(0,2)$ tensor multiplet may take near the location of the defect. We can write our general solution in \eqref{1by16BPSsolutionform} as
  \begin{align}
  		F(\zeta \Phi_0, \zeta \Phi_1, \zeta \Phi_2, \zeta \Phi_3) = G( \zeta \Phi_0, \Phi_1/\Phi_0, \Phi_2/\Phi_0, \Phi_3/\Phi_0 ) = 0
  \end{align}
  When $G$ is a linear function of $\zeta \Phi_0$ we can write
  \begin{align}
  		\label{GofZPhi}
  		&G( \zeta \Phi_0, \Phi_1/\Phi_0, \Phi_2/\Phi_0, \Phi_3/\Phi_0 ) \cr 
  		& \quad = \zeta \Phi_0 \, F_1( \Phi_1/\Phi_0, \Phi_2/\Phi_0, \Phi_3/\Phi_0 ) - F_0(  \Phi_1/\Phi_0, \Phi_2/\Phi_0, \Phi_3/\Phi_0 ) = 0 \cr
  \end{align}
  and very near the boundary we have
  \begin{align}
  		\label{profilefrombulk}
  		\lambda \nu_0 = \frac{F_0( \nu_1/\nu_0, \nu_2/\nu_0, \nu_3/\nu_0 )}{F_1(  \nu_1/\nu_0, \nu_2/\nu_0, \nu_3/\nu_0 )} \,.
  \end{align}
  Here $\lambda = r e^{i \frac{\xi_1}2}$ is a field on the probe brane that determines the radial and angular profile of the probe brane and may be identified with two of the real scalars in the 6d boundary $(2,0)$ tensor multiplet theory. The defects on the boundary are given by zero-sets of $F_1(  \nu_1/\nu_0, \nu_2/\nu_0, \nu_3/\nu_0 )$. And therefore, our analysis suggests that in \eqref{profilefrombulk} is the singularity profile that the boundary theory scalars can take near the defect. We can consider the half-BPS solution in the previous section \ref{halfBPSM5no1} as an example. The holomorphic function $F$ in that case will be $F(\zeta \Phi_0, \zeta \Phi_1, \zeta \Phi_2, \zeta \Phi_3)$ = $\zeta \Phi_1 - c$ and very near the boundary we have the following expression for $\lambda$
  \begin{align}
  		\label{profilehalfBPS}
  		\lambda = \frac c{\nu_1}  =  \frac c{\cos \alpha \, e^{i \phi_1}}\,.
  \end{align}
  Therefore, near the AdS boundary region where $\alpha \rightarrow \frac{\pi}2$ we have the singular behaviour for $\lambda$ due to the simple pole. \\
  
  \noindent
   To summarize, in this section, we have analyzed the M5 brane solutions that wrap a one-dimensional circle in the $S^4$ part of the geometry. We have taken the near $AdS$ boundary limit for the non-compact brane solutions from which our analysis suggests the location of the general $\frac 1{16}$-BPS codimension-two defects in the 6d boundary $(2,0)$ gauge theory, described by the equation \eqref{codimension2location}. Our analysis also suggests that how in the presence of such defects, some of the scalar field components may acquire a singular behaviour near their location: for a half-BPS static solution, from \eqref{profilehalfBPS}; for a general $\frac 1{16}$-BPS abelian solution, from \eqref{profilefrombulk}. For further understanding of these codimension-two defect solutions in the boundary gauge theory more work is required in the gauge theory side and we hope to report more about them in future work. \\
   
  	\vspace{.4cm}
   \noindent
   \textbf{\large Acknowledgements} \\
   
   We are grateful to Sujay Ashok for his support, for many discussions throughout this work, and comments on this manuscript. We thank Nemani V. Suryanarayana for the correspondence during the initial phases and valuable comments on the work appeared here. We like to thank K. Narayan for related discussions when this work was in progress. The author is grateful to the Institute of Mathematical Sciences, where he had an affiliation as a graduate student.
   
   
  
  \providecommand{\href}[2]{#2}\begingroup\raggedright\endgroup

\end{document}